\documentclass[twocolumn,trackchanges]{aastex63}
\usepackage{graphicx}	
\usepackage{amsmath}
\usepackage{amssymb}	
\usepackage{newtxtext,newtxmath}
\usepackage{xcolor}
\usepackage[T1]{fontenc}
\usepackage{ae,aecompl} 
\usepackage{tikz}

\let\tablenum\relax
\usepackage{siunitx}

\newcommand\aastex{AAS\TeX}
\newcommand\latex{La\TeX}

\received{Sep 28, 2021}
\revised{\today}
\accepted{\today}
\shorttitle{{\it XMM-Newton} and {\it NuSTAR} observations of PSR~J1653$-$0158}
\shortauthors{Long et al}
\graphicspath{{./}{figures/}}

\begin{document}

\title{{\it XMM-Newton} and {\it NuSTAR} observations 
 of the  compact millisecond pulsar binary PSR~J1653$-$0158}

\correspondingauthor{Jane SiNan Long (JSL), Albert K. H. kong (AKHK), Kinwah Wu (KW)}
\email{sinan$\_$jane@gapp.nthu.edu.tw (JSL), akong@phys.nthu.edu.tw (AKHK), 
kinwah.wu@ucl.ac.uk (KW)}

\author[0000-0002-2126-0050]{Jane SiNan Long}
\affiliation{
Institute of Astronomy, National Tsing Hua University, Hsinchu 30013, 
  Taiwan (ROC)} 
\author[0000-0002-5105-344X]{Albert K. H. Kong}
\affiliation{
Institute of Astronomy, National Tsing Hua University, Hsinchu 30013, 
  Taiwan (ROC)} 
\author[0000-0002-7568-8765]{Kinwah Wu}
\affiliation{
Mullard Space Science Laboratory, University College London, 
  Holmbury St Mary, Surrey, RH5 6NT, United Kingdom}
\author{Jumpei Takata}
\affiliation{
School of Physics, Huazhong University of Science and Technology, Wuhan, China}
\author[0000-0002-2342-9956]{Qin Han}
\affiliation{
School of Astronomy and Space Science, Nanjing University, Nanjing, 210023, China}
\affiliation{
Mullard Space Science Laboratory, University College London, 
  Holmbury St Mary, Surrey, RH5 6NT, United Kingdom}
\author[0000-0003-1753-1660]{David C. Y. Hui}
\affiliation{
Department of Astronomy and Space Science, Chungnam National University, Daejeon, Korea (ROK)}
\author[0000-0002-0439-7047]{Kwan Lok Li}
\affiliation{
Department of Physics, National Cheng Kung University, 
Tainan, Taiwan (ROC)}



\begin{abstract}
We have presented the first joint {\it XMM-Newton} and {\it NuSTAR} analysis
  of the millisecond pulsar (MSP) binary PSR J1653$-$0158. 
The 75-minute orbital period inferred from   optical and gamma-ray observations  
  together with the 1.97-ms pulsation in the gamma-rays 
  indicate that this system is 
  the most compact Black Widow MSP system known to date. 
The orbital period was not detected 
  in the {\it XMM-Newton} and {\it NuSTAR} data, 
  probably due to insufficient photon counts obtained in the observations.     
Fitting the joint X-ray spectrum of PSR J1653$-$0158 with a power law 
  gives a photon index $\Gamma = 1.71 \pm 0.09$.    
The X-ray luminosity of the source in the ($0.2-40$)~keV band 
 is deduced to be $1.18 \times 10^{31}\;\!{\rm erg~s}^{-1}$,  
 for an adopted distance of 0.84~kpc.  
We have shown that the broad-band X-ray spectrum
  can be explained by synchrotron radiation from electrons  
  accelerated in the intra-binary shock,   
  and the gamma-rays detected in the {\it Fermi} data are curvature radiations from 
 electrons and positrons in the pulsar magnetosphere.
    Our kinematic analysis 
 of the Tidarren systems PSR~J1653--0158 and PSR~J1311--3430 
   indicates that the two Tidarren systems are likely to have originated in the Galactic Disk.
\end{abstract} 



\section{Introduction}
\label{sec:introduction}  

The {\it Fermi}-LAT source 4FGL J1653.6$-$0158 (= PSR J1653 $-$0158) 
  was proposed as a gamma-ray emitting millisecond pulsar (MSP) binary,   
  when a variable X-ray and optical source with a 75-min periodicity was found within the gamma-ray positional uncertainty \citep{Kong2014ApJL,Romani2014ApJL}. 
The subsequent detection of a 1.97-ms pulsation in the gamma-ray band
  confirmed its nature as a MSP 
  \citep{Nieder2020ApJL}. 
Compact MSP binary systems with binary periods 
  as short as that of PSR~J1653$-$0158 
  would have a low-mass  semi-degenerate companion 
  \citep[see e.g.][]{Bhattacharya1991PhR,Iben1997ApJ,Chen2013ApJ,Jia2014ApJ,Hui2018ApJ}.  
The deduced mass of $\sim 0.014~{\rm M}_{\odot}$ \citep{Nieder2020ApJL} 
of the companion
is above the critical mass limit $\sim 0.006~{\rm M}_{\odot}$ 
  for dynamically stable mass transfer \citep[see e.g.][]{Kiel2013}.  
With continuous ablation 
  by the energetic particles 
  and evaporation by the radiation 
  from the MSP, 
  the companion star may
   lose all its mass completely,   
  leaving only an isolated MSP in the system  
  \citep{Kluzniak1988Nat,Phinney1988Nat,Ruderman1989ApJ,Faucher2006ApJ}.
  
Compact MSP binaries 
  exhibit two distinctive observational behaviors, 
  by which they are classified into two groups, 
  with names assigned after two spider families:  
  the Redback (RB) and the Black Widow (BW) 
  \citep[see e.g.][]{Chen2013ApJ, Roberts2013Proc}.   
RBs are believed to be systems 
  in the transition from/between accretion and rotation-powered phases. Their companions are mostly partially degenerate stars that are filling the Roche lobe 
  or very close to filling the Roche lobe. The 
BW systems are characterised by the ablation of the  highly degenerate companion.  
They are not powered by the accretion processes 
  and therefore are not X-ray luminous.   
Although some  compact MSP binaries can switch 
  between being rotation powered and accretion powered 
  \citep[see e.g.][]{Papitto2013Nat}, 
  depending on the relative sizes of the companion stars 
  and their Roche lobes,   
  they would eventually 
  become persistently rotation powered.  
These systems would resemble the BW systems if the   companions fail to regain contact with their critical Roche surfaces.  
The currently known  
  RB generally have companions with mass $M_{\rm c} \gtrsim 0.1~{\rm M}_{\odot}$ 
  \citep[see e.g.][]{Hui2019}.  
   compact MSP binaries 
 with   companion mass $M_{\rm c} \lesssim 0.05~{\rm M}_{\odot}$ 
  almost certainly belong to the BW group 
  \citep[see][]{Fruchter1988Nat, Stappers1996ApJL}.  
Some studies \citep[e.g.][]{Chen2013ApJ} 
  suggest that most RBs and BWs 
  are descendants of different groups of systems, 
  implying that the most of observed RB are unlikely to have evolved from the BW, 
  despite that RB can switch off accretion permanently.

The detection of the millisecond gamma-ray pulsations 
  in PSR~J1653$-$0158 
  implies that the MSP is presently not accreting.  
The low luminosity of the X-rays, 
  about $10^{31}\;\!{\rm erg~s}^{-1}$,   
  observed in the source \citep[][]{Kong2014ApJL} 
  is consistent with no significant  
  mass transfer within the system.   
This, together with the deduced low   companion mass \citep{Nieder2020ApJL},  
  readily puts PSR~J1653$-$0158 as a BW, 
  with its pulsar emissions powered
  by the extraction of the rotational energy of the neutron star.

This paper reports the findings 
  from a joint multi-wavelength 
  timing and spectral analysis 
  of the {\it XMM-Newton}, {\it NuSTAR} and {\it Fermi} observations 
  of PSR~J1653$-$0158. 
\S2 presents the observational set-ups 
 and \S3 reports the temporal and spectral analyses. 
\S4 discusses the results from the analysis.  
We adopted an intra-binary shock model to explain the observed broadband X-ray 
  spectral properties, as well as a magnetosphere model \citep{Takata2012ApJ} to explain the gamma-ray spectral behavior. 
The origin of PSR~J1653$-$0158  
  and its related compact MSP binaries 
  are also discussed.  

\section{observations} 
\label{sec:observation}

\subsection{NuSTAR} 
\label{subsec:NuSTAR}

PSR~J1653$-$0158 was observed
  by {\it NuSTAR} \citep{Harrison2013ApJ} on May 29, 2017 
 for about 102~ks (ObsID 30201017002; PI: Kong). 
The data were processed with the {\it NuSTAR} Data Analysis Software 
  NUSTARDAS (v1.9.6),  
  using the calibration data from CALDB version 20200813.
  Procedures with standard parameters in the {\it NuSTAR} Data Analysis Software Guide \footnote{\url{https://heasarc.gsfc.nasa.gov/docs/nustar/analysis/nustar_swguide.pdf}}  
  were adopted to clean and filter the event lists. 
  The calibrated and cleaned event lists
  were processed with the tool {\tt nupipeline} following standard procedures.  
The HEASoft tool {\tt nuproducts} 
  were used to construct the response matrices for each 
  of the two focal plane modules, FPMA/B,  
  and to produce images, light curves, spectra of the source. 
 The FPMA/B net counts were $\sim$ 118 counts and $\sim$ 78 counts, respectively.

In our analysis, 
 the energy range was set to be 3$-$40~keV 
 as there were almost no source photons above 40 keV.  
Images, light curves, and spectra of the target
  were derived from the data extracted from a circular region 
  with a radius of 20 arcsec centered at the X-ray position of PSR~J1653$-$0158. 
An annulus region with a width of 40 arsec and an inner radius of 20 arcsec 
  centered at the source 
  were used to derive the background photons. 
The spectra of the source from FPMA and FPMB observations 
   were rebinned such that there were at least 10 counts in each spectral bin.

\subsection{XMM-Newton} 
\label{subsec:XMMs} 

PSR~J1653$-$0158 was observed by {\it XMM-Newton} on March 09, 2017 
  (ObsID: 0790660101; PI: Kong). 
  The total exposure time was 53~ks, 
  with data obtained from the EPIC (European Photon Imaging Camera)  
  MOS1, MOS2 and pn CCD detectors.  We followed the data analysis procedure detailed in the Data Analysis Threads Version 7.0 provided by SAS v19.0 \footnote{\url{https://heasarc.gsfc.nasa.gov/docs/xmm/abc}}. Due to the very low signal-to-noise ratio of MOS1 and MOS2 data, we only used pn data in this work.
The pn camera has a good sensitivity below 3~keV, 
  which compensates the lack of sensitivity of {\it NuSTAR} 
  in low energies, 
  hence provides an essential constraint  
  for the soft X-rays in spectral analysis.  
The raw data (ODF) were processed 
  to be used with {\tt xmmextractor} together with calibration data provided by the Current Calibration Files (CCF). 
The pn event lists were further processed 
  with the EPIC reduction meta-tasks {\tt emproc} and {\tt epproc}, respectively. To filter the EPIC event lists for flaring background, 
   the 10$-$12 keV light curve was examined 
   with the SAS tool {\tt evselct} and filtered the flare 
   by setting Rate Expression 'RATE<=0.4'. The effective exposure time is $\sim 25$ ks after background flaring filtering. The cleaned event lists were then used 
  to produce the light curves and spectrum. The radius of the source is chosen as $\sim$ 10 arcsec, and an annulus region with a width of 30 arsec and an inner radius of 15 arcsec 
  centered at the source 
  were used to derive the background photons. 
After background subtraction, the   net counts for pn are $\sim$ 893 counts.

\subsection{Fermi-{\rm LAT}} 
\label{subsec:NuSTAR} 

The {\it Fermi}-LAT data (the latest version, P8R3) 
  from August 04, 2008 to March 19, 2021 (spanning over 150 months) 
  were analysed using the {\tt Fermitools}. 
The on-source data was extracted from a region of $20^{\circ}$ radius 
 centred at the 4FGL~J1653.6$-$0158 position,  
 (RA, Dec) $=(253^{\circ}.408, -1^{\circ}.97667)$,  
 with energies between 100 MeV and 300 GeV. 
The tracker in the front and back sections of all the events were included, 
  from which we selected {\tt evtype = 3} and filtered the data 
  with an event class {\tt evtclass = 128 } assuming PSR~J1653$-$0158 (= 4FGL~J1653.6$-$0158) as a point source. 
To avoid the gamma-ray contamination coming from the Earth's albedo, 
  photons with zenith angles smaller than $90^{\circ}$ were selected.   
Furthermore, the selection was restricted to 
 high quality data in the time intervals (i.e. choosing DATA\_QUAL>0). Binned likelihood analysis was performed 
  using the {\it Fermi} science tool {\tt gtlike}.
To eliminate the background distribution, a background emission model, 
  which included 
  the Galactic diffuse emissions ({\tt gll\_iem\_v07.fits}) 
  and the isotropic diffuse emissions ({\tt iso\_P8R3\_SOURCE\_V3\_v01.fits}) given by the {\it Fermi} Science Support Center, 
  was applied. To obtain the best-fitting spectral model for 4FGL~J1653.6$-$0158, we applied the user contributed tool {\tt make4FGLxml.py} that uses the spectral model from the 4FGL catalog \citep{Abdollahi2020} to calculate the flux contribution of each source in the $20^{\circ}$ radius region of interest centered at 4FGL~ J1653.6$-$0158 position. The TS value obtained by the source model {\tt PLSuperExpCutoff2} in 100 MeV to 300 GeV energies is 4643.61.
  

\section{Temporal and spectral analysis} 
\label{sec:analysis_results}

\subsection{Temporal Behavior} 
\label{subsec:timing}

PSR J1653$-$0158 has an orbital   period of 0.0519~d shown in the optical \citep{Kong2014ApJL,Romani2014ApJL}, gamma-ray \citep{Nieder2020ApJL}, and possibly X-ray \citep{Kong2014ApJL} 
  wavebands. 
In a previous X-ray study, the 75-min orbital period found in optical is marginally shown in the {\it Chandra} data \citep{Kong2014ApJL}.  By using the larger collecting area of {\it XMM-Newton}, we investigated the X-ray modulation in detail. We also used {\it NuSTAR} to investigate the light curve in the hard X-ray region.

\begin{figure}[ht!]
\vspace*{0.5cm}
\epsscale{1.1}
\plotone{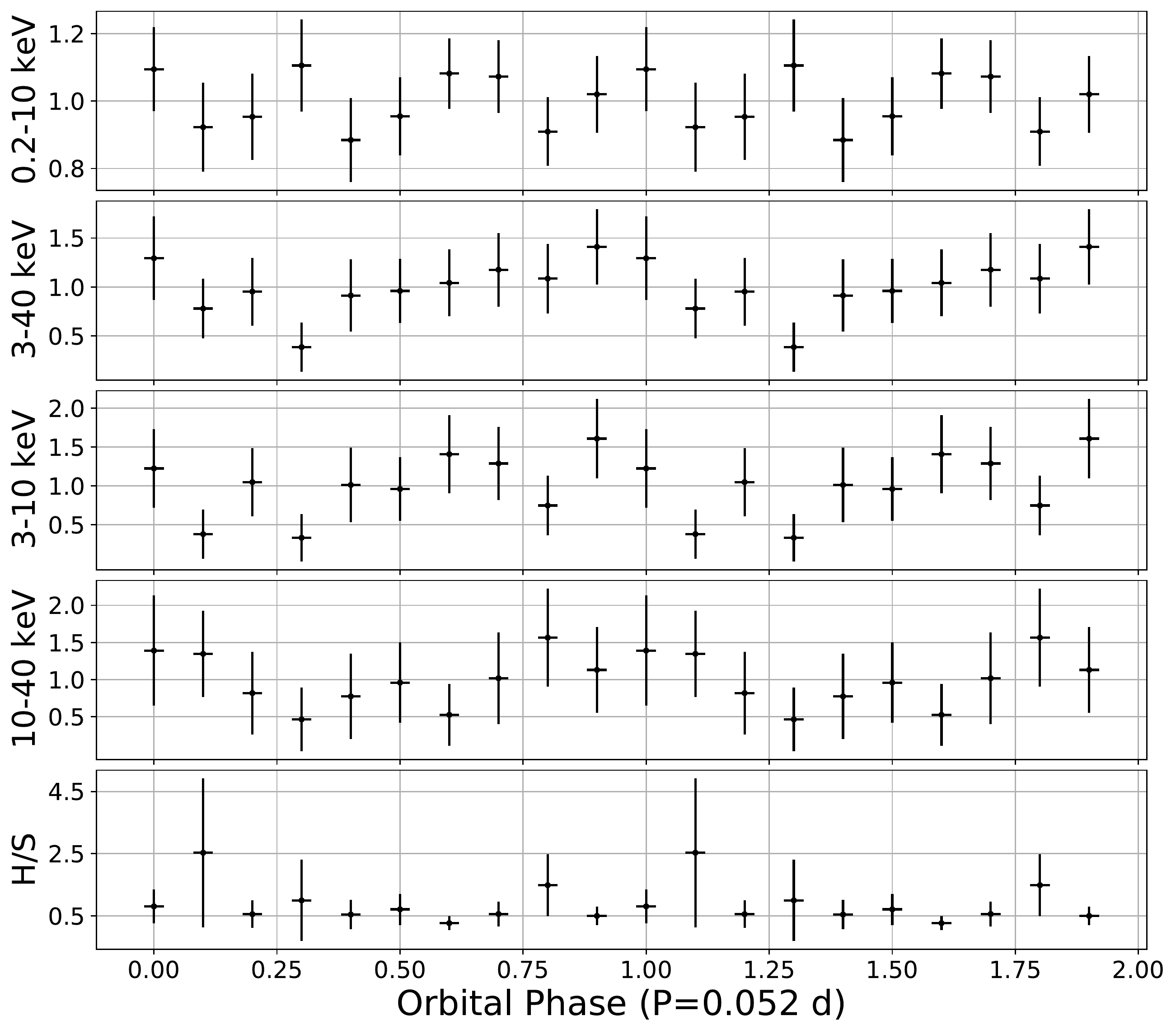}
\caption{The folded light curves, in normalised counts, of PSR J1653--0158 
from {\it XMM-Newton} (0.2-10 keV) and {\it NuSTAR} (3-40 keV) observations. 
 Orbital cycles with different energy bands and hardness ratio ($10-40$~keV/$3-10$~keV) are shown for clarity. 
The light curves do not show the 75-min orbital period.}
\label{fig:lc}
\end{figure}

Fig.~\ref{fig:lc} shows 
 the {\it XMM-Newton} folded light curves 
 in the $0.2-10$~keV band with the gamma-ray epoch $T_{\rm asc}=$ MJD 56513.479171(8) and an    orbital period of 0.0519447575(4)~d   measured from  {\it Fermi} gamma-ray observations \citep{Nieder2020ApJL}.  
To show the broadband X-ray variability, we also plotted the {\it NuSTAR} folded light curve 
 ($3-40$~keV).   Furthermore, we plotted the hardness ratio (H/S) between hard X-rays ($10-40$~keV) and soft X-rays ($3-10$~keV). We found no evidence for the 75-min orbital modulation in either {\it XMM-Newton} or {\it NuSTAR} data.

The modulations in the light curves were assessed 
  using the Lomb-Scargle periodogram 
  \citep{Lomb1976,Scargle1982}.
For the {\it XMM-Newton} normalized light curve, 
  the Lomb-Scargle power at the 75-min orbital period 
  corresponds to a 99\% false alarm probability.  
A 3-$\sigma$ upper limit of 24\% for the amplitude 
  was obtained by fitting a sinusoidal function of 75-min period. 
The false alarm probability is 3.7$\%$ 
  and the  3-$\sigma$ upper limit for the amplitude is 80\% for the {\it NuSTAR} light curve.
The hardness ratio light curve is shown in Fig~\ref{fig:lc},  
  for completeness, 
  and it does not show evidence of spectral variations.  

\subsection{Spectral Properties} 
\label{subsec:spectral}

\subsubsection{X-ray}

We used XSPEC version 12.11 to perform X-ray spectral fitting. Since both the {\it NuSTAR} and {\it XMM-Newton} observations were taken at similar epochs in 2017, we fitted the energy spectra from {\it XMM-Newton}, {\it NuSTAR}'s FPMA and FPMB observations simultaneously to increase the signal-to-noise ratio. We also performed some simple spectral fits of individual spectra and they show no significant flux and spectral changes.

\begin{figure}[ht!]
\vspace*{0.5cm}
\epsscale{1.1}
\plotone{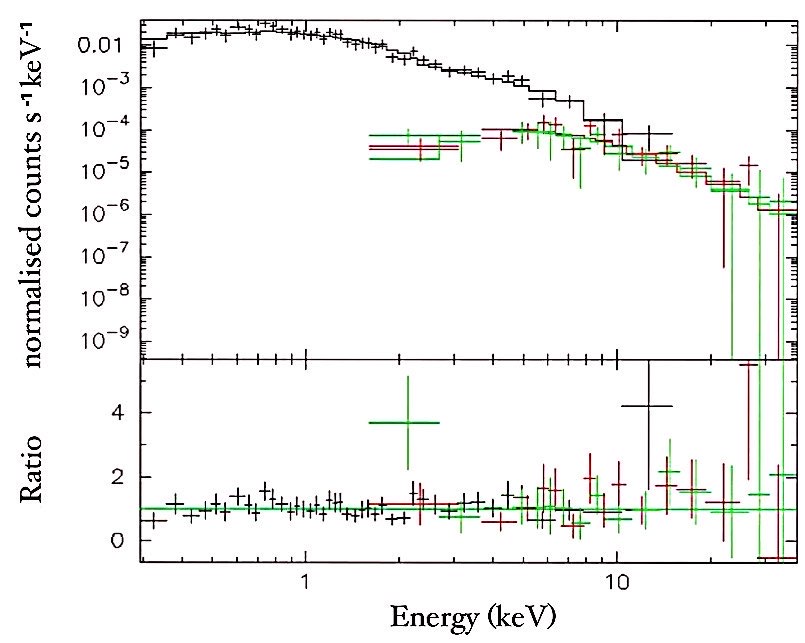}
\caption{Model fit to the{\it XMM-Newton} ($0.2-10$ keV) and {\it NuSTAR} ($3-40$ keV) spectra 
 of PSR~J1653$-$0158. The best-fitting model is an absorbed power-law with 
 a photon index $\Gamma =1.71$ (shown as the dark solid line). }
\label{fig:spectrum}
\end{figure}

We employed different in-built models in XSPEC to perform spectral fitting. Based on previous X-ray study \citep{Hui2015ApJ}, we first tried an absorbed simple power-law model. In order to fit the spectra from the three cameras (pn and MOS1/2) of {\it XMM-Newton} and the two cameras of {\it NuSTAR} simultaneously,  cross-calibration factors were taken into account in all the spectral models. In general, the 0.2$-$40 keV X-ray emissions can be well described with an absorbed power-law model ($\chi^2 =51.35$ for 66 degrees of freedom (dof)) without obvious emission and absorption features 
(Fig.~\ref{fig:spectrum}). 
The best-fit absorption value is 
 $(8.85\pm2.29) \times 10^{20}~{\rm cm}^{-2}$,  consistent with the extinction $A_V = 1.06$ obtained from light curve modelling \citep{Nieder2020ApJL}, 
 while the best-fit photon index 
is $1.71\pm0.09$. The unabsorbed $0.2-40$ keV flux is
 $1.40^{+0.13}_{-0.12} \times  10^{-13}~{\rm erg\;\!cm}^{-2}{\rm s}^{-1}$, 
  corresponding to an X-ray luminosity of
  $1.18 \times 10^{31}{\rm erg\;\! s}^{-1}$ 
    at a distance of 0.84 kpc from optical modeling 
   \citep{Nieder2020ApJL}.  

Although an absorbed power-law model can provide a reasonable best-fit, 
we also investigated if neutron star thermal emission from PSR~J1653$-$0158 contributes part of the X-ray emissions \citep[e.g.][]{Kong2018MNRAS}. 
We included a non-magnetic neutron star atmosphere component 
in the absorbed power-law model \citep[nsatmos model in XSPEC;][]{Heinke2006ApJ}. 
We fixed the mass of the neutron star to be  $2.17~{\rm M}_{\odot}$ \citep{Nieder2020ApJL}. 
Without losing generality, 
 we adopted a value of 10~km as the radius of the neutron star
 \citep[see][]{Lattimer2001ApJ,Abbott2018PhRvL}.  
  The effective temperature derived from the model is 
   $7.99^{+4.02}_{-2.68} \times 10^5~{\rm K}$ and the unabsorbed flux in $0.2-40$ keV is 
   $6.27^{+0.35}_{-0.33} \times  10^{-14}~{\rm erg\;\!cm}^{-2}{\rm s}^{-1}$,  
   corresponding to a luminosity of
   $5.29 \times 10^{30}{\rm erg\;\! s}^{-1}$    for a distance of 0.84 kpc.
The best-fitting parameters from both spectral models 
  are presented in Table~\ref{fit-par}. We applied a likelihood ratio test to test the validity of an extra component. A ratio of 0.988 suggests that a simple power-law model is sufficient. Furthermore, we used F-test to investigate if the additional neutron star atmosphere component is significant. The F-test probability is 0.0186 indicating that the additional component is not statistically required.

\subsubsection{Gamma-ray}

\begin{figure}[ht!]
\vspace*{0.5cm}
\epsscale{1.1}
\plotone{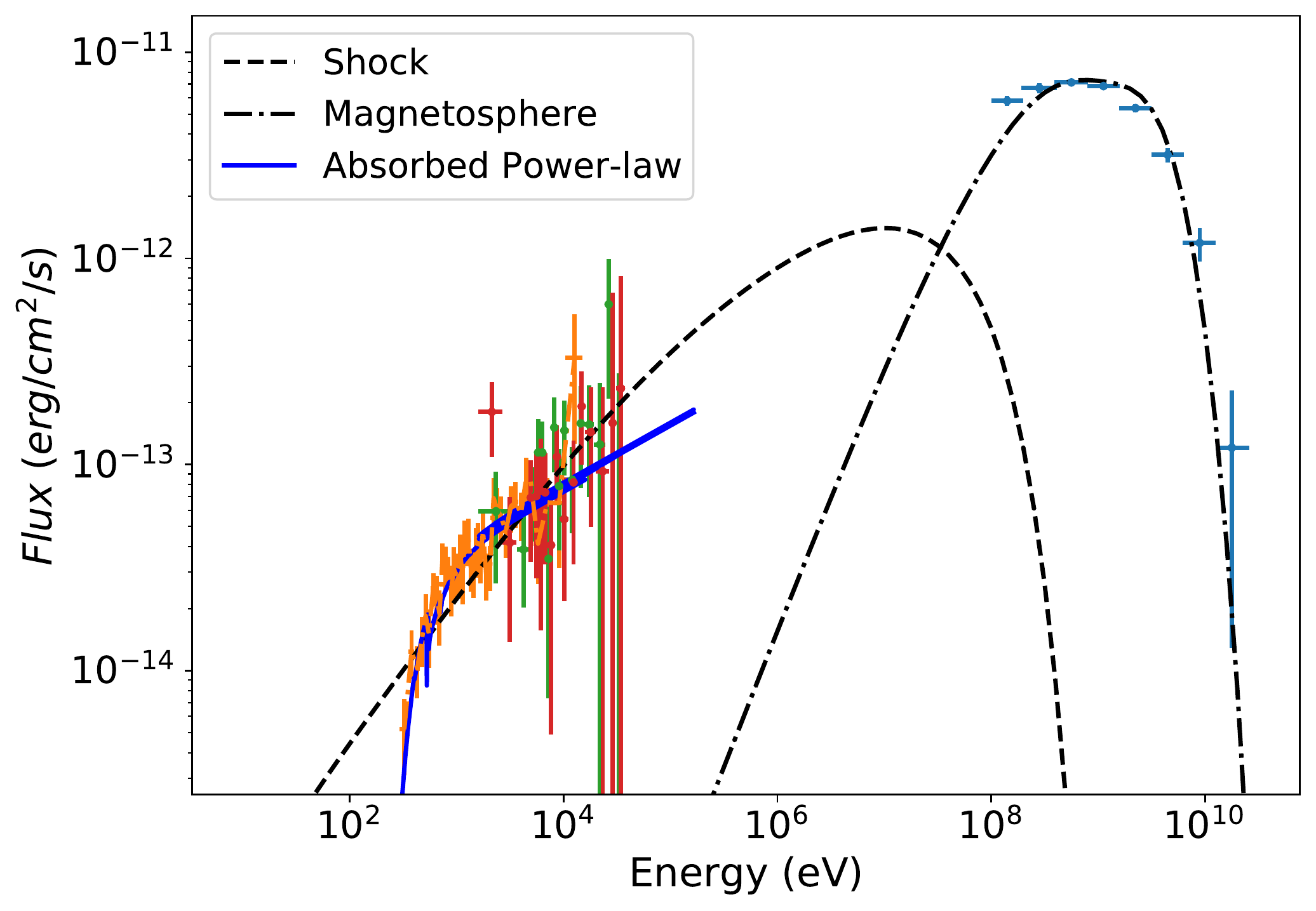}
\caption{Broad-band spectrum, in X-ray and gamma-ray energies,  
 of PSR~J1653$-$0158 fitted with an intra-binary shock 
 and magnetosphere model 
 \citep{Takata2012ApJ}. 
 We assumed a momentum ratio of $\eta_b=0.7$, 
  a shocked pulsar wind velocity of $v \sim 0.25~c$, and an inclination of $60^{\circ}$. 
The best-fit absorbed power-law model for the X-ray spectrum is also shown (thick blue curve).}
\label{fit-model}
\end{figure}

For the GeV band, we divided the {\it Fermi}-LAT photon counts data into 8 energy segments to obtain the gamma-ray spectrum (see blue crosses in Fig~\ref{fit-model}). We fitted the gamma-ray spectrum of PSR J1653$-$0158 with a power-law and an exponential cut-off model:
\begin{equation}
\frac{\text{d} N}{\text{d}E} = N_0 \left( \frac{E}{E_0} \right)^{\Gamma} \text{exp} (- a E^b ) ,
\end{equation}
where $N$ is the photon counts per unit time, unit area and $E$ is the photon energy, $N_0$ and $E_0$ are the normalization factors, $\Gamma$ is the spectral index and $a$ is the exponential factor. By setting 
$b= {2}/{3}$ 
\citep[an empirical value chosen for pulsars, see][]{Abdollahi2020}, 
  $a =  (8.4 \pm 0.74) \times 10^{-3} \ {\rm MeV}^{- \frac{2}{3}}$ 
and $\Gamma = 1.58 \pm 0.05$ were obtained. 
Note that the total flux is $F = (5.06 \pm 0.19) \times 10^{-8} \text{ photons cm}^{-2} \text{s}^{-1}$.

\begin{table}
    \begin{center}
    \caption{Spectral fits for PSR J1653$-$0158. Fluxes $F$ 
   are from combined {\it XMM-Newton} and {\it NuSTAR} unabsorbed    flux   and a distance of 0.84~kpc is assumed in calculations.  \label{fit-par} }
    \begin{tabular}{lcr}
    \hline 
    Model & $\qquad \qquad \qquad \quad  $ & \\
    \hline
    Power-law & $\qquad$ &  \\
     $N_{\rm H} (10^{20} \text{cm}^{-2})$    & $\qquad$ & $8.85\pm2.29$ \\
        $\Gamma$ & $\qquad$ & $1.71\pm0.09$ \\
       $F_{0.2-40} (10^{-13} \text{erg cm}^{-2} \text{s}^{-1})$  & $\qquad$ &  $1.40^{+0.13}_{-0.12}$\\
       $\chi_{\nu}^2/ \text{dof}$  & $\qquad$ & 0.78/66 \\
         & $\qquad$ &  \\
       Power-law + H atmosphere$^a$  & $\qquad$ &  \\
        $N_{\rm H} (10^{20} \text{cm}^{-2})$ & $\qquad$ & $24.38\pm14.31$ \\
$\Gamma$ &  & $1.60\pm0.15$  \\
$T(10^5 \text{K})$&  & $7.99^{+4.02}_{-2.68}$ \\ 
$F_{0.2-40} (10^{-14} \text{erg cm}^{-2} \text{s}^{-1})$ &  & $6.27^{+0.35}_{-0.33}$ \\
$\chi_{\nu}^2/ \text{dof}$ &  & 0.71/64 \\ 
\hline
    \end{tabular}
    \end{center}
\tablecomments{$^a$ The mass and radius of the neutron star were fixed to be 
     $2.17\;\!\text{M}_{\odot}$ \citep{Nieder2020ApJL} and $10~{\rm km}$, 
     assuming a distance of 0.84 kpc.}
\end{table}  

\section{Discussion} 
\label{sec:discussion}

\subsection{Theoretical Interpretation for the High-energy Emissions}
\label{subsec:model fitting}


\begin{deluxetable*}{cccccccccc}
\tablecaption{Physical Parameters of PSR~J1653$-$0158 and the other Tidarren systems
\label{table1}}
\tablewidth{0pt}
\tablehead{
\colhead{Source} & \colhead{$D$} &  \colhead{$P$} &
 \colhead{$L_X$} & \colhead{$L_{\rm sd}$} & \colhead{$P_{\rm orb}$} & \colhead{$M_{\rm com}$} & \colhead{$M_{\text{NS}}$ $^f$}   & \colhead{DM }\\
\colhead{} & \colhead{(kpc)}  &  \colhead{(ms)} &
\colhead{($10^{31}$ erg ${\rm s^{-1}}$)} &  \colhead{($10^{33}$ erg ${\rm s^{-1}}$)} & \colhead{(day)} & \colhead{($M_{\odot}$)} &
\colhead{($M_{\odot}$)}    & \colhead{(pc $\text{cm}^{-3}$)}}
\startdata
PSR J1653$-$0158 & 0.84$^a$ &  1.97 & 1.18 & 4.4 & 0.052 & 0.013 & 1.62 &  $-$ \\
PSR J0636+5129 & 0.5$^b$ &  2.8 & 4.48$^c$ & 5.6 & 0.066 & 0.0068 & $-$   &   11.1\\
PSR J1311$-$3430 & 1.4$^d$ & 2.56 & 5.6 & 49 & 0.065 & 0.011 & 1.53  &  37.8
\enddata
\tablecomments{$^a$ The distance of $840 \pm 40$~pc is obtained from optical modelling \citep{Nieder2020ApJL}. \\
$^b$ The distance of 0.5 kpc is derived from dispersion measurement \citep{Stovall2014}. \\
$^c$ The X-ray luminosity is calculated from power-law modelled flux  $15^{+2}_{-7} \times  10^{-13} \ \text{erg cm}^{-2}  \ \text{s}^{-1}$ \citep{Spiewak2016}. \\
$^d$ The distance of 1.4 kpc is derived from dispersion measurement \citep{Ray2013}.   \\
$^e$   The minimum pulsar mass is estimated from binary mass function and companion radial velocity amplitude $K=666.9 \ \text{km s}^{-1}$(J1653$-$0158),  $K=609.5 \ \text{km s}^{-1}$(J1311$-$3430). 
J0636+5129 is lack of optical radial velocity information \citep{Draghis2018, Kaplan2018, Spiewak2018}.}

\end{deluxetable*} 



\begin{deluxetable*}{ccccccccc}
\tablecaption{Location and kinetics of PSR~J1653$-$0158 and the other Tidarren systems  
\label{table2}}
\tablewidth{0pt}
\tablehead{
\colhead{Source} & \colhead{$R_{\rm C}$} & \colhead{$l$} & \colhead{$b$} & \colhead{$z$} &   \colhead{$\mu_{\alpha}$} & \colhead{$\mu_{\delta}$} & \colhead{$V_{r}$} & \colhead{Ref}\\
\colhead{} & \colhead{(kpc)} &   \colhead{(deg)} &  \colhead{(deg)} &
\colhead{(kpc)} &  \colhead{($\text{mas yr}^{-1}$)} & \colhead{($\text{mas yr}^{-1}$)} & \colhead{($\text{km s}^{-1}$)}  & \colhead{} }
\startdata
PSR J1653$-$0158 & 7.28 & 16.61  & 24.93 & 0.36 &  $-19.62 \pm 1.86$ & $-3.74 \pm 1.12$ &  $-174.6 \pm 5.1$ & $a$  \\
PSR J0636+5129 & 8.46 & 163.91 & 18.64 & 0.16 &  $3.22 \pm 0.03$ & $-1.61 \pm 0.06$ &   $-$ & $b$ \\
PSR J1311$-$3430 & 6.89 &307.68&28.17& 0.79 &  $-6.8 \pm 0.6$ & $-3.5 \pm 0.8$ 
  & $62.5 \pm 4.5$ & $c$ 
\enddata
\tablecomments{
The references are: 
$a$. \cite{Romani2014ApJL, Nieder2020ApJL}; 
$b$. \cite{Stovall2014, Guillot2019}; 
$c$. \cite{An2017, Romani2012} \\ 
The definitions of the symbols are $R_{\rm C}$: Distance to Galactic center, adopting the distance from the Sun to the Galactic centre $R_0 = 8~{\rm kpc}$ \citep{Camarillo2018PASP}; $l$:  Galactic longitude; $b$: Galactic latitude; $z$: Distance to the Galactic plane; $\mu_{\alpha}, \mu_{\delta}$: Proper motions in right ascension and declination; $V_{r}$: mean radial velocity of the binary system.} 
\end{deluxetable*} 


We present an analysis  
  of the broad-band ($0.2-40$~keV) X-ray data    
  of the  compact BW PSR J1653$-$0158 
  obtained by {\it XMM-Newton} and {\it NuSTAR}. 
While the {\it Chandra} observation  
  in the $0.3-8$~keV energy band
  indicated a possible period of about $75$ min 
 \citep{Kong2014ApJL}, 
 we found no clear modulation 
 in the {\it XMM-Newton} and {\it NuSTAR} data. 
The null detection could be  
  due to poorer photon statistics,  
 as the point spread functions of {\it XMM-Newton} and {\it NuSTAR} are much broader than that of {\it Chandra}. 
The background contribution of the {\it XMM-Newton} and {\it NuSTAR} 
  light curves
  is 27\% and $52-65$\%, respectively.
  On the other hand, the background contribution 
  is negligible (almost 0$\%$) in 
  the {\it Chandra} light curve. 

Although a composite model,  
  consisting of a power-law and a neutron star atmosphere component, 
  fits the X-ray spectrum (up to about 40~keV) 
  of PSR~J1653$-$0158 well,  
  a single component absorbed power-law model 
  is sufficient. 
The parameters of 
 $\Gamma = 1.71\pm0.09$ 
 and $N_{\rm H}=(8.86\pm2.29) \times 10^{20}{\rm cm}^{-2}$ 
 obtained from the absorbed power-law fit 
 is consistent with those obtained in the previous analysis 
 of {\it Chandra} observation   
 \citep{Kong2014ApJL,Romani2014ApJL}. 
The photon index of $\Gamma = 1.71 \pm 0.09$ 
 implies a spectral index $\alpha = 0.71\pm 0.09$ 
 (as $\alpha=\Gamma-1$). 
If the X-rays are optically thin synchrotron radiation 
 from non-thermal relativistic electrons 
 with a power-law energy distribution,  
 in a uniformed magnetised medium, 
 we expect the power-law index of the electrons  
 to be $p\approx2.4$ (as $\alpha = (p-1)/2$)
 \citep[see e.g.][]{Rybicki1986book}. 
Stochastic accelerations in shocks 
  can produce energetic electrons with
  a power-law energy distribution ($p \approx 2.2-2.5$), 
  in both relativistic and non- relativistic regime
  \citep[see e.g.][]{Bell1978MNRAS,Achterberg2001MNRAS}. 
If the X-rays from PSR~J1653$-$0158 
  is of synchrotron origin, 
  then they could be emitted
  from energetic electrons 
  accelerated in the bow shock 
  formed when pulsar wind 
  collided with stellar material
  ablated from the companion star.   The existence of the ablating wind can also be inferred by optical observations, as indicated by the decreasing modulation and flat orbital minima in the blue colours. Optical light curve modeling must include a non-thermal veiling flux component which could be explained by synchrotron emission from an intra-binary shock \citep{Romani2014ApJL, Nieder2020ApJL}.

 
We considered an intra-binary shock 
  and a magnetosphere model \citep{Takata2012ApJ}  
   to explain the general spectral behavior in the X-ray and gamma ray bands.   
This intra-binary model 
  was previously applied 
  to explain the broad-band high-energy spectrum
  of the RB system PSR J2129$-$0429 \citep{Kong2018MNRAS}, 
  which has a non-degenerate companion star.   
In PSR J2129$-$0429, 
  the intra-binary shock has 
  a momentum ratio of $\eta_{\rm b} \approx 7$  
  (where $\eta_{\rm b}$ is the ratio  
  between the stellar magnetic pressure 
  and the ram pressure of the pulsar wind).   
As the stellar wind dominated the flow, 
  the intra-binary shock wrapped around the pulsar. 
In this study, we considered that the intra-binary shock 
  in PSR J1653$-$0158 
  was produced by the collision 
  of an isotropic pulsar wind 
  with an envelope of material ablated 
  from a white-dwarf companion. 
The intra-binary shock 
  accelerated the electrons and positrons 
  to relativistic energies and they  
  emitted synchrotron X-rays. 
Different to PSR J2129$-$0429,
  the intra-binary shock 
  in PSR~J1653$-$0158
  was located closer to the white-dwarf companion 
  and wrapped around it. 
We adopted a magnetization parameter   
  $\sigma = 0.1$,  
  for the ratio of the magnetic energy 
  and kinetic energy of the pulsar wind, 
  and a momentum ratio $\eta_{\rm b} \approx 0.7$  
  in the model to fit the X-ray spectra 
  of PSR J1653$-$0158.     
The pulsar wind carried out the spin-down power 
  and was compressed by the shock.  
The shock provided a mean to accelerate 
  the charged electrons 
  to relativistic energies, 
  which emitted the synchrotron X-rays.   
Fig~\ref{fit-model} 
  shows the X-ray intra-binary model 
 (dashed line) fit to the observed broad-band X-ray spectrum. 
 



The observed gamma-rays are not emitted 
  from the high-energy electrons associated with the shock 
  but are instead produced by 
  the energetic charged particles in the pulsar magnetosphere 
  \citep{Cheng1986, Dyks2003, Watters2009}.
They are curvature radiation 
  from relativistic electrons and positrons  
  created through pair processes
  in the pulsar magnetosphere.  
This scenario provides an explanation 
  to the double-peak features 
  observed in the gamma-ray light curves 
  of MSP binaries
 \citep[e.g.][]{Huang2012ApJ,Li2014ApJ}.  

We applied a three-dimensional two-layer outer 
  gap model of \cite{Wang2011} to 
  calculate the gamma-ray spectrum 
  of PSR J1653$-$0158. 
We estimated,  
  from the spin period and the surface magnetic field, 
  that the thickness of the outer gap 
  is about 60\% of the light cylinder radius.  
This indicates 
  that the large fraction of the volume  
  in the outer magnetosphere 
  is occupied by the outer gap. 
We considered  
  that the outer gap exists 
  between the null charge surface of the Goldreich-Julian charge density 
  and the light cylinder.  
This would produce a pulse profile 
  consistent with the broad pulse profile 
  as observed 
  and also the high-efficiency 
  ($L_{\gamma}/L_{\rm sd} \sim 66\%$) 
  in the GeV energies \citep[see][]{Nieder2020ApJL}. 
We assumed a value for the electric current 
  corresponding 
  to $\approx 50\%$ of Goldreich-Julian density  
  and  calculated the electric field 
  along the magnetic field line. 
Fig.~\ref{fit-model} 
  shows the spectrum of the 
  curvature gamma-rays produced by our model\footnote{GeV 
   gamma-rays can be produced in the pulsar magnetosphere 
 when low-energy photons are Compton up-scattered by 
 relativistic electrons and positrons  
 \citep[see e.g.][]{Grenier2015CRPhy}.},   together with the {\it Fermi} data.

\subsection{Origin of     PSR~J1653$-$0158}
\label{subsec:origin}  
\begin{figure}[ht!]
\vspace*{0.5cm}
\epsscale{1.25}
\plotone{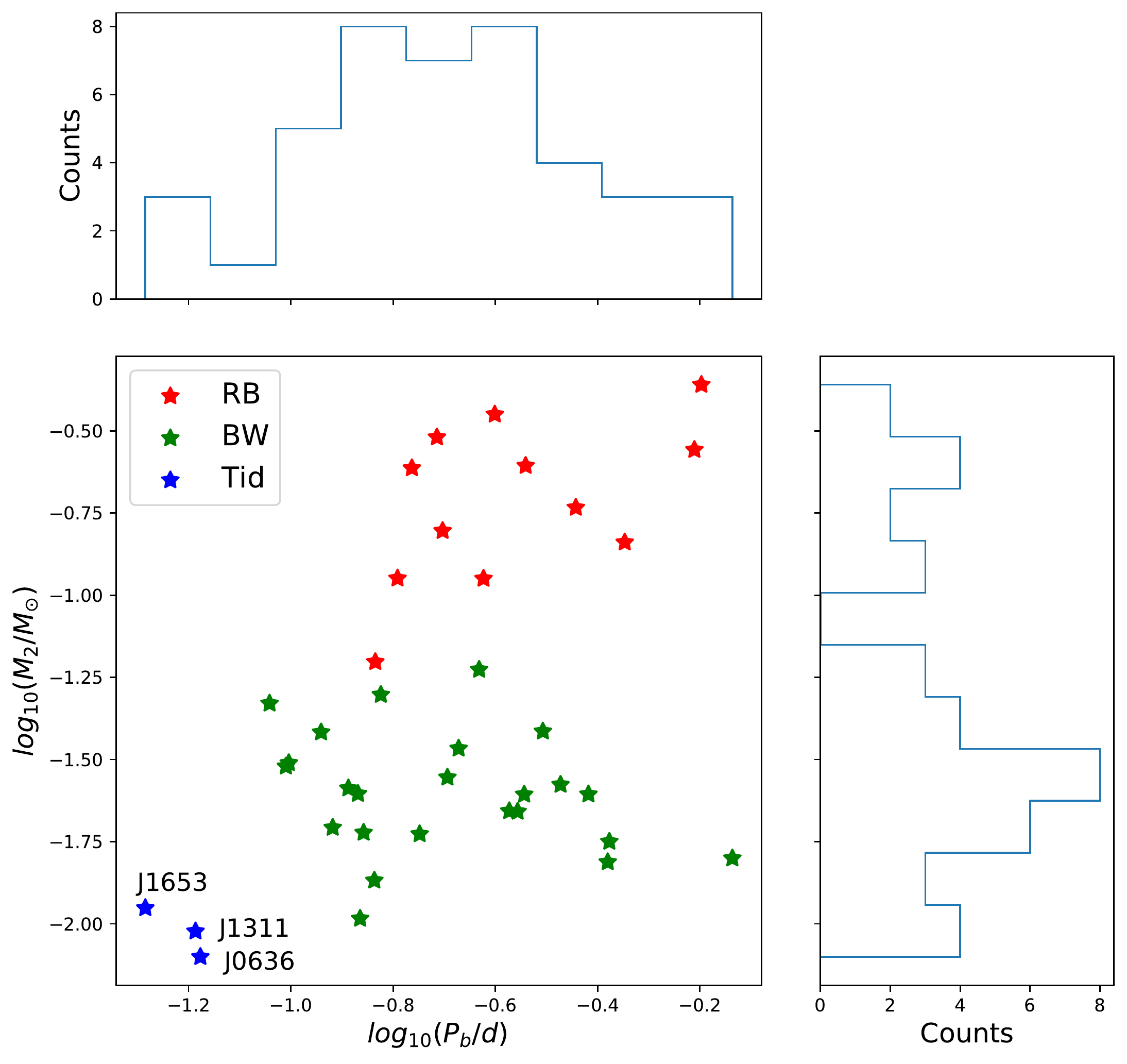}
\caption{The distribution of eclipsing MSP binaries in the Galactic field 
 are presented in the orbital period - companion mass plane. 
The BWs and RBs occupy two separate regions 
  distinguished by the companion mass, 
  while they are statistically indistinguishable by orbital period. 
The Tidarrens occupy the the region 
  visually well distinguishable from those resided by the RBs and BWs, 
  and they are separated from the other two groups 
  by both the companion mass and the orbital period. 
The Tiderrens PSR~J1653$-$0158, PSR~J1311$-$3430 and PSR~J0636+5129 are mark in blue and labelled respectively. } 

\label{group}
\end{figure}

 We showed all spider MSP systems which has the characteristic eclipsing light curve in the Galactic field in Fig.\ref{group}. The RBs and BWs are distinguished by their companion mass, whereas the subclass Tidarren (Tid) is distinguished from the main BW class by their companion mass as well as orbital period. The $p$-values resulting from the two-sample Kolmogorov-Smirnov (KS) tests between Tid and BW are $2.7 \times 10^{-3}$ for mass and $5.4 \times 10^{-4}$ for period, indicating the differences between the two classes.
   
PSR~J1653$-$0158 
and two other compact MSP systems,   
PSR~J0636+5129 and PSR~J1311$-$3430
 \citep[see][]{Draghis2018,vanHaaften2012,Romani2015, Spiewak2018}, 
 are known as the Tidarren 
 \citep[][]{Romani2016Tid}.
As a subclass of the BW systems,  they have
 a very low mass companion star 
  and extremely short orbital period 
 \citep[][]{Romani2016Tid}, 
 and their properties are shown in Table \ref{table1}. The companion stars in the Tidarren systems  
  have strongly heated sides facing the pulsar. 
This leads to periodic variations
  in the optical emissions of the system,     
  providing us a mean to derive the orbital velocities 
  of the companion stars \citep[see e.g.][]{Draghis2019, Kandel2019}.  
The companion stars of the Tidarren systems 
  have an extremely low mass.    
Their hydrogen is almost completely stripped, 
  and hence they often appear as helium WDs. The Tidarren systems are therefore more likely to be the progenitors 
  of isolated MSPs than the other subclass of MSP binaries.   

 The formation of eclipsing MSP systems in the Galactic field is thought to undergo the recycled process similar to the evolution of CV-like LMXBs \citep{Chen2013ApJ, Ginzburg2021}. The bimodal distribution of the RBs and BWs can be explained by different evaporation efficiency. However, none of the evolution tracks can match the observed quantities of the Tidarren systems. A different formation mechanism is proposed by \cite{King2003, King2005} that a MSP-WD binary is originally formed in the globular clusters (GC) and exchange its companion to a main-sequence star and subsequently ejected to the Galactic field or entered the field populations when their host GCs dissociated \citep{Gnedin1997}. Therefore, We assessed the possibilities of Tidarrens origin
 by comparing the trajectories of the two systems  
 with the distributions of binaries populations 
 from the Galactic Disk or GCs.


All the known Tidarren systems are located at substantial Galactic latitude 
 (with $|b| > 12~{\rm deg}$). 
From their measured distances to Earth, 
  we determined their vertical distances to the Galactic plane $z$. The
$z$ of all Tidarren systems are larger than the scale height of the Galactic Thin Disk ($\sim$ 0.12~kpc), 
  and  two of them, PSR J1653$-$0158, PSR J1311$-$3430, 
  have $z$ larger than the scale height of the Galactic Thick Disk 
  \citep[$\sim$ 0.3~kpc;][]{de_Jong2010, Juri2008}. 
Adopting the distance from the Sun to the Galactic centre 
  $R_0 = 8~{\rm kpc}$ \citep[see][]{Eisenhauer2003,Francis2014MNRAS,Vall2017,Camarillo2018PASP, Evgeny2021}, 
  we derived the distances of all known Tidarren systems to the Galactic center $R_{\rm C}$ in Table \ref{table2}. 
The values of their $R_{\rm C}$ are about 
 $6.6 - 8.5~{\rm kpc}$, larger than $2~{\rm kpc}$, 
  the radius of the Galactic bulge \citep[see][]{Zoccali2016PASA}. 
We therefore conclude that 
 the currently known Tidarren systems
 are not in the Galactic bulge or in the Galactic Thin Disk.  
A possible explanation for the spatial locations 
of the Tidarren systems   
  is that they originated from globular clusters (GCs).   
To examine the scenario that the Tidarren systems 
  were produced in GC, 
  we first compared the population of BWs and binary MSP in GC and in the field. 
The current version of ATNF 
Pulsar Catalogue\footnote{\url{http://www.atnf.csiro.au/research/pulsar/psrcat}}, 
  listed 64 GC binary MSPs and 163 field binary MSPs, 
  and a recent study by \cite{Hui2019} listed 17 BWs in GC and 29 BWs in field. 
This gives a ratio of 0.26 for BWs among binary MSPs in GC 
  and 0.16 for BWs among binary MSPs in the field, 
  consistent with that BWs have no preference to reside in a GC
  \citep[cf.][]{King2003}.

\begin{figure}[ht!] 
\vspace*{0.5cm}
\epsscale{1.15}
\plotone{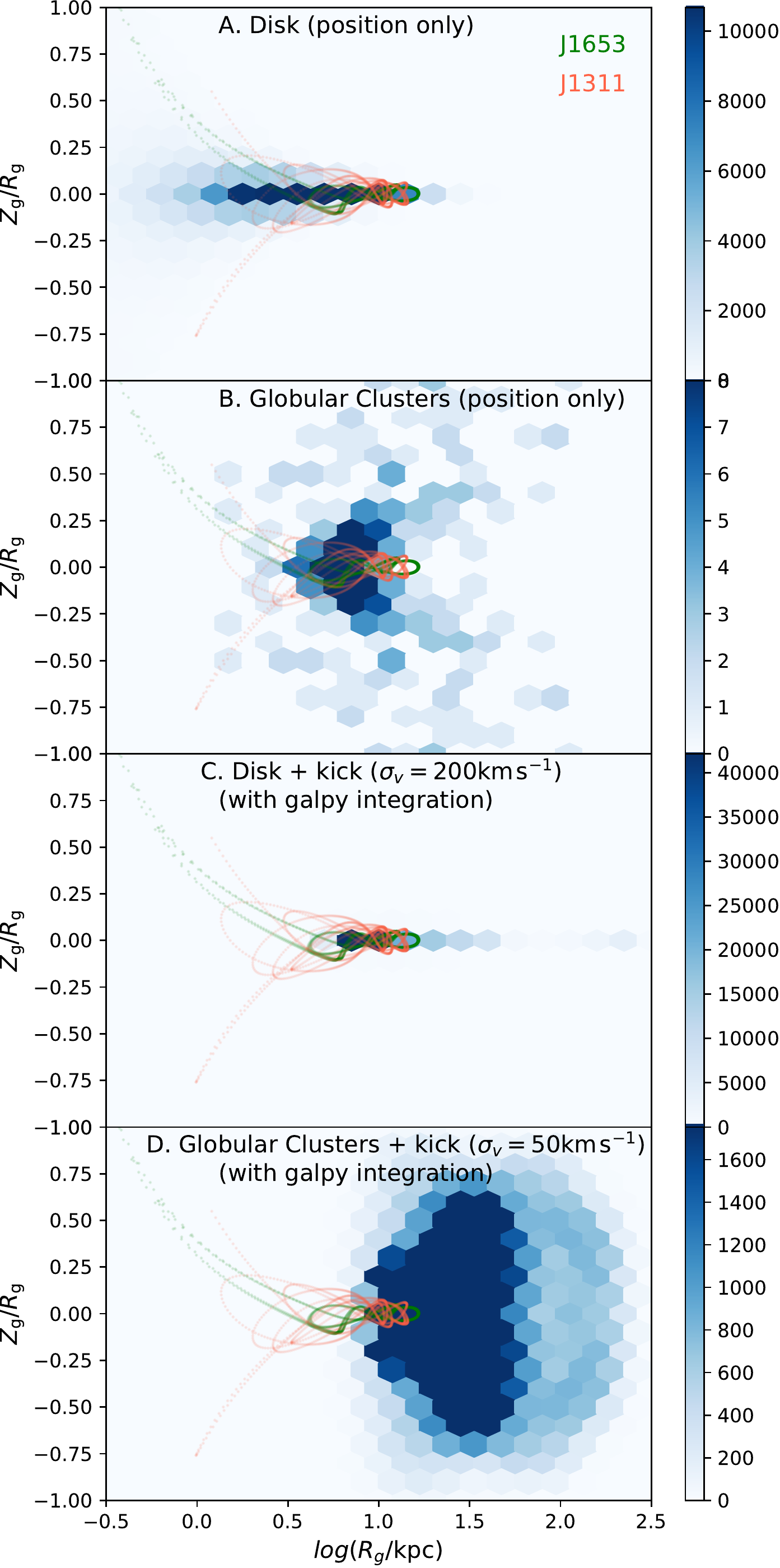}
\caption{  The orbits 
  of PSR~J1653$-$0158 
  and PSR~J1311$-$3430  
  in the past 1~Gyr, overlaid with the distributions of binary population from Galactic Disk or globular clusters (obtained from catalog or via Monte Carlo simulation). 
The x and y axes represent 
 the radial distance to Galactic Center (in $\log {R_{\rm g}}$)  
 and the polar angle (in $Z_g/R_{\rm g}$) respectively. 
In panels A and C,
the positions of binary population from the Galactic Thin Disk is sampled with a scale height of 0.12~kpc~\citep{Rix2013} and a scale length of 4.0~kpc~\citep{de_Jong2010}, 
  and in panels B and D, from GCs listed in the \cite{Harris1996} catalogue 
  (2010 edition). 
Systems in panels A and B do not receive a kick at birth. 
Systems in panels C and D received a kick, 
  which has an isotropic Maxwellian velocity distribution 
  with $\sigma_v=200~{\rm km~s}^{-1}$ 
  and $50~{\rm km~s}^{-1}$ respectively. 
}
\label{fig:GC_bl}
\end{figure}









Among the three systems listed in Table \ref{table2}, 
  PSR~J1653$-$0158 and PSR~J1311$-$3430 
  have both mean radial velocity and proper motion measurements. 
The orbits of these systems in the Milky Way can therefore be computed.  
We used \texttt{galpy} \citep{Bovy2015}\footnote{\texttt{galpy} can be downloaded from \url{http://github.com/jobovy/galpy}.} to track back the orbit of them in the past 1~Gyr. 
  
   Fig.~\ref{fig:GC_bl} shows the orbits of PSR~J1653$-$0158 and PSR~J1311$-$3430 and binary populations from Galactic Thin Disk or GC on the $\log{R_{\rm g}}$-$Z_{\rm g}/R_{\rm g}$ plane, where $R_{\rm g}$ is the radial distance 
    from the Galactic Centre, 
    $Z_{\rm g}$ is the z component of  Galactocentric Cartesian coordinate 
    and  $Z_{\rm g}/R_{\rm g} = \cos \theta$ 
    where $\theta$ is the polar angle. 
The time-averaged absolute values 
  of the Galactic latitude ($\langle \vert b\vert \rangle_t$)  
  is $2.5^{\circ}$ for PSR~J1653$-$0158 
  and $3.7^{\circ}$ for PSR~J1311$-$3430.  
This gives the time-averaged absolute distances $\langle \vert z\vert \rangle_t$)  of $0.3~{\rm kpc}$ to the Galactic plane for PSR~J1653$-$0158 
  and of $0.45~{\rm kpc}$ for PSR ~1311$-$3430.      
The time-averaged distances to the Galactic centre  
  is   $9.7~{\rm kpc}$ for PSR~J1653$-$0158 
  and   $8.4~{\rm kpc}$ for PSR~J1311$-$3430. As their distances to the Galactic Centre 
 are larger than $2~{\rm kpc}$,  
 PSR~J1653$-$0158 and PSR~J1311$-$3430  
are unlikely associated with the Galactic bulge stellar population.   The binary populations in Fig.~\ref{fig:GC_bl} are obtained as follows. 
We used Monte Carlo (MC) methods to sample the binary populations from the Galactic Thin Disk, 
  where the thin disk assumes 
  a scale height of 0.12~kpc~\citep{Rix2013} 
  and a scale length of 4.0~kpc~\citep{de_Jong2010},  
  in the panel A of Fig.~\ref{fig:GC_bl}. 
The GC populations in panel B was read directly 
 from the \cite{Harris1996} catalogue (2010 edition). 
Only position information of these populations are shown in panels A and B, and no orbital integration were preformed.

 For panels C and D, we made further assumptions about the initial velocities  for these binaries populations, and also about the kick velocities they received. We sampled $N=10^5$ systems from the Galactic Thin Disk and $N=10^5$ systems from 
 GCs and performed orbital integration of those systems and calculated their time-averaged $R_{\rm g}$ and $Z_{\rm g}$. 

For panel C, the binary population from the Galactic Thin Disk, their initial velocities on the plane before kicks were calculated following the rotation curve from \citep{Sofue2017}, 
  and the vertical velocity was assumed to be 0. 
The Maxwellian distribution of the kick velocity is characterised  
  by $\sigma_v=200~{\rm km~s}^{-1}$,  
  appropriate for the kick received by the binary   
  in the supernova explosion that produced the neutron stars. 

For the binary population from GCs in panel D, the initial 3D velocity 
follows a Maxwellian distribution (d.o.f=3). 
The parameter which determines the distribution was calculated by $\sqrt{3}a_1$, where $a_1=75.33~{\rm km~s}^{-1}$ 
is the parameter obtained by fitting the radial velocities of GCs with 
a Maxwellian distribution (d.o.f=1). 
We added small kicks with velocities 
following a Maxwellian distribution with 
$\sigma_v=50~{\rm km~s}^{-1}$, 
 corresponding to the recoil velocity of the system   
 when leaving the GC. 
All the kick velocities are isotropic  
(evenly distributed over 4$\pi$ solid angle) in the rest frame of the binaries. 

The trajectories of PSR~J1653$-$0158 and PSR~J1311$-$3430 
   tend to coincide with 
    systems of Galactic Disk origins 
   rather than systems of GC origins. 
For the cases with position information only 
  (without orbital integration), 
  the trajectories of the two systems 
  are consistent with systems  
  of Galactic Disk and GC populations. 
When the kinetic of the systems of the two populations 
  are properly accounted for, 
  the trajectories of PSR~J1653$-$0158 and PSR~J1311$-$3430 
  are consistent with the expectations from the systems 
  associated with the Galactic Disk but 
  inconsistent with the systems associated with GCs.   
This can be understood as follows.  
The velocities of the population of systems 
  from the Galactic Disk are jointly determined 
  by their rotational motion around the Galactic centre 
  and their kick velocity. 
Among the two velocities, 
  the rotation component 
  do not affect the time-averaged positions of the systems 
  during orbital integration, and 
  it also dilutes the effects 
  brought by the kick velocity. 
For the population of systems from  GCs, 
  the kick velocity is relatively small, 
  and the movements were determined 
  by the (3D) velocities of GCs. 
In our calculations, 
  the velocities of GCs 
  were derived from radial velocities 
  provided by the GC catalogue, 
  and  a significant fraction of the systems have relatively large radial velocities in Galactocentric coordinate.  
This introduces substantial scatters  
  towards larger $R_{\rm g}$ in the distribution, 
  which is at odd to the expected locations of PSR~J1653$-$0158 and PSR~J1311$-$3430 
  from their computed past trajectories. 
In summary, 
  our kinematic analyses 
  have shown that 
  PSR~J1653$-$0158 and PSR~J1311$-$3430  
  are more likely to have originated  from  
  the Galactic Disk rather than GCs.
 

\section{Conclusion} 

We presented a broadband timing and spectral analysis of 
 the BW MSP binary PSR J1653$-$0158 using {\it XMM-Newton}, {\it NuSTAR}, and {\it Fermi}-LAT data.
Our analysis did not reveal  
  detectable periodic modulations in the 
  0.2$-$40~keV energy band. 
The null detection
  of the binary orbital modulation 
  could be due to 
  the substantial background contribution 
  to the photon counts.  
We found that the X-ray spectrum 
  can be modelled by an absorbed power-law, 
  with the best-fit photon index $\Gamma = 1.71\pm0.09$, 
  (spectral index $\alpha = 0.71\pm0.09$), 
  a value typical  
  of optically thin synchrotron radiation 
  from electrons freshly accelerated in shocks 
  via stochastic processes.   
The unabsorbed X-ray flux, up to 40~keV  
 was determined to be  
 $1.40^{+0.13}_{-0.12} \times  
   10^{-13}~{\rm erg\;\!cm}^{-2}{\rm s}^{-1}$,
  implying an X-ray luminosity of
  $1.18 \times 10^{31}{\rm erg\;\! s}^{-1}$ 
   for a distance of   0.84~kpc 
   derived from the optical observations.
We examined if the X-ray emission  
  would be contributed by the pulsar atmospheric emissions. 
The addition of a neutron star atmosphere 
component to the absorbed power-law spectrum  
  gave a photon index of $\Gamma = 1.60\pm0.15$. 
However, this additional spectral component is not statistically significant.
 
We modeled the combined X-ray and gamma-ray spectra 
  of PSR J1653$-$0158 with an intra-binary shock 
  and magnetospheric emission model. 
The intra-binary shock is formed 
  when the pulsar wind collides 
  with the media ablated 
  from the semi-degenerate companion, 
  and the synchrotron X-rays 
  are emitted from the electrons 
  accelerated by the shock. 
The gamma-rays are produced by curvature radiation from  
  energetic charged particles 
  in the pulsar magnetosphere. 
  
The origin of PSR J1653$-$0158 
  and its similar systems were discussed. 
We grouped the BW systems 
  with extremely low mass companion star 
  and extremely short orbital period
  as the Tidarren systems 
  and conducted an analysis 
  of their spatial location in the Milky Way 
  and their kinetic properties. 
We found that  
   these Tidarren systems
   have radial distances $R_{\rm C} \sim 6.6 - 8.5~{\rm kpc}$ 
   to the Galactic center 
   and vertical distances 
    $\ z \sim 0.16 - 0.79~{\rm kpc}$ 
   to the Galactic plane,   
   implying that they 
   are not currently located in the Galactic bulge 
   or the Galactic Thin Disk.

The possibilities 
  that the two Tidarren systems originated from Galactic Disk or GCs 
  were assessed 
  using their computed trajectories from {\tt galpy} in the past 1~Gyr.  
Their trajectories indicated that  
 PSR~J1653$-$0158 and PSR~J1311$-$3430 
 have been residing within 
 a radial distance of $\sim 16~{\rm kpc}$  
 from the Galactic Centre.  
PSR~J1653$-$0158  
  had a time-averaged distance 
  of   $9.7~{\rm kpc}$ to the Galactic centre  
  and a time-averaged distance   
   of $0.3~{\rm kpc}$ from the Galactic plane.  
PSR~J1311$-$3430 
   had a time-averaged distance 
   of   $8.4~{\rm kpc}$ to the Galactic centre 
   and a time-averaged distance of 
   of $0.45~{\rm kpc}$ from the Galactic plane. 
We further conducted 
  a more detailed kinematic analysis 
  of the populations of similar compact binaries, 
  assuming origins from the Galactic Disk and GCs. 
Comparing their distributions 
  with the computed past trajectories 
  of PSR~J1653$-$0158 and PSR~J1311$-$3430 
  suggests that the two Tidarren systems 
  are likely mkred to have originated in  the Galactic Disk.


\section*{Acknowledgements} 
We thank the referee for the critical comments 
  and useful suggestions 
  that led to substantial improvement 
  of the science of this work. 
This work is supported in part by the Ministry of Science and Technology 
  of Taiwan (ROC) under the grants 109-2628-M-007-005-RSP and 110-2628-M-007-005 
  (PI: A.~Kong). 
KW is supported in part by a UK STFC Consolidated Grant 
 to UCL-MSSL. 
JT acknowledges the support by National Key 
   R\&D Program of China, 
   2020YFC2201400, NSFC U1838102. 
CYH is supported by the National Research Foundation of Korea 
   through grants 2016R1A5A1013277 and 2019R1F1A1062071.  
KLL is supported by the Ministry of Science and Technology 
   of Taiwan (ROC) 
  through grant 109-2636-M-006-017, 
  and by the Ministry of Education of Taiwan (ROC) 
  through a Yushan (Young) Scholarship.  
Analyses conducted 
  by QH at UCL-MSSL were supported by a 
  UCL Overseas Research Scholarship 
  and a UK STFC PhD Studentship.
This work has made use of NASA's Astrophysics Data System 
  and the CSIRO ATNF Pulsar Catalogue \citep{Manchester2005}. The updated list of eclipsing spider MSP binaries in the Galactic field is provided by Jane Yap.

\software{XSPEC (v12.11; \cite{Arnaud1996}), SAS (v19.0; \cite{Gabriel2004}), NUSTARDAS (v1.9.6; \cite{Harrison2013}), Fermitools (Fermi Science Support Development Team 2019), make4FGLxml.py, galpy \citep{Bovy2015}} 


 
\bibliography{J1653}{}
\bibliographystyle{aasjournal}



\end{document}